\documentclass[a4paper]{PoS}
\usepackage[square,sort,comma,numbers]{natbib}
\usepackage{aas_macros}

\usepackage{soul}
\sethlcolor{yellow}

\title{Longitudinal and transverse velocity fields in parsec-scale jets}

\ShortTitle{Longitudinal and transverse velocity fields in parsec-scale jets}

\author{\speaker{Florent Mertens}$^1$ and Andrei Lobanov$^{1,2}$\\
        \llap{$^1$}Max-Planck-Institut f\"ur Radioastronomie\\
        Auf dem Hugel 69, 53121 Bonn, Germany\\
        \llap{$^2$}Institut f\"ur Experimentalphysik, Universit\"at Hamburg\\
        Luruper Chaussee 149, 22761 Hamburg, Germany\\
        E-mail: \email{fmertens@mpifr.de}, \email{alobanov@mpifr.de}}

\abstract{Radio-loud AGN typically manifest powerful relativistic jets
extending up to millions of light years and often showing superluminal motions
organised in a complex kinematic pattern. A number of physical models are still
competing to explain the jet structure and kinematics revealed by radio images
using the VLBI technique. Robust measurements of longitudinal and transverse
velocity field in the jets would provide crucial information for these models.
This is a difficult task, particularly for transversely resolved jets in objects
like 3C\,273 and M\,87. To address this task, we have developed a new technique for
identifying significant structural patterns (SSP) of smooth, transversely
resolved flows and obtaining a velocity field from cross-correlation of these
regions in multi-epoch observations. Detection of individual SSP is performed
using the wavelet decomposition and multiscale segmentation of the observed
structure. The cross-correlation algorithm combines structural information on
different scales of the wavelet decomposition, providing a robust and reliable
identification of related SSP in multi -epoch images. The algorithm enables
recovering structural evolution on scales down to 0.25 full width at half
maximum (FWHM) of the image point spread function (PSF). We
present here examples of applying this algorithm to obtain the
first detailed transverse velocity fields and to study the kinematic evolution
in the parsec-scale jets in 3C\,273 and M\,87. }

\FullConference{12th European VLBI Network Symposium and Users Meeting,\\
    7-10 October 2014\\
    Cagliari, Italy}

\begin{document}

\section{Introduction}

The steady improvements of dynamic range of astronomical images and
increasing complexity and detail of astrophysical modeling bring a
higher demand on automatic (or {\em unsupervised}) methods for
characterization and analysis of structural patterns in astronomical
images.

A number of approaches developed in the fields of computer vision and remote
sensing for tracking structural changes \citep[{\em   cf.},]
[]{zitova_image_2003} typically
require {\em oversampling} in the temporal domain. This renders them
difficult to be used in astronomical applications that requires
tracking changes monitored with sparse
sampling and dealing with partially transparent, {\em optically thin}
structures.

Presently, structural decomposition of astronomical images normally
involves simplified {\em supervised} techniques based on
identification of specific features of the structure \citep{lobanov_a_1998} or fitting the observed
structure with a set of predefined templates ({\em e.g.},
two-dimensional Gaussian features).

A more robust approach to automatize identification and tracking of structural
patterns in astronomical images can be provided by a generic multiscale method
such as wavelet deconvolution or wavelet decomposition \citep[{\em
cf.},][]{starck_astronomical_2006}. In \citep{mertens_waveletbased_2014}, we
explored this wavelet approach and present a wavelet-based image segmentation
and evaluation (WISE) method for structure decomposition and tracking in
astronomical images. In this contribution, we present an overview of this new
method along with results obtain on WISE analysis of parsec-scale radio jets in
3C\,273 and 3C\,120 compared with results of conventional structure analysis
previously applied to these data. Finally, preliminary results of the velocity
field analysis of the jet in M\,87 is presented.

\section{Wavelet-based Image Segmentation and Evaluation}

To characterize the structure and structural evolution of an
astronomical object, the imaged object structure needs to be
decomposed into a set of significant structural patterns (SSP) that
can be successfully tracked across a sequence of images. The
multiscale decomposition provided by the wavelet transform
\citep{mallat_a_1989} makes wavelets exceptionally well-suited to
perform such a decomposition, yielding an accurate assessment of the
noise variation across the image and warranting a robust
representation of the characteristic structural patterns of the
image. To adapt better to the needs of astronomical imaging, we have
extended the multiscale approach to object detection, similarly to the
methodology developed for the multiscale vision model
\citep[MVM;][]{starck_astronomical_2006}. The method employs segmented wavelet decomposition (SWD) of
individual images into arbitrary two- dimensional SSP (or image
regions) and subsequent multiscale cross-correlation (MCC) of the
resulting sets of SSP.

\subsection{Segmented Wavelet Decomposition}

The segmented wavelet decomposition (SWD) comprises the following steps
to describe an image structure by a set of statistically
significant patterns:

\begin{itemize}
\itemsep0em 
\item[1.]~A wavelet transform is performed on an image $I$  by
decomposing the image into a set of $J$ sub-bands (scales), $w_{j}$,
  and estimating the residual image noise (variable across the image).

\item[2.]~At each sub-band, statistically significant wavelet
  coefficients are extracted from the decomposition by thresholding
  them against the image noise.

\item[3.]~The significant coefficients are examined for local maxima,
  and a subset of the local maxima satisfying detection
  criteria is identified, defining the locations of SSP in
  the image.

\item[4.]~Two-dimensional boundaries of the SSP are defined 
  by the watershed segmentation using the feature
  locations as initial markers.
\end{itemize} 


The decomposition provides a structure
representation that is sensitive to compact and marginally resolved features as
well as to structural patterns much larger than the FWHM of the instrumental PSF
in the image.

\subsection{Multiscale Cross-Correlation}

To detect structural differences between two images of an
astronomical object made at epochs $t_1$ and $t_2$, one needs to find
an optimal set of displacements of the original SSP (described by the
groups of SSP $S_{j,1}, j=1,...,J$) that would match the SSP in
the second image (described by $S_{j,2}$,
$j=1,...,J$). Cross-correlating $S_{j,1}$ and $S_{j,2}$ is a
well-suited tool for this purpose. There are two specific issues
that need to be addressed, however, to ensure that  the
cross-correlation analysis is reliable. First, a viable rule needs to be
introduced to identify the relevant image area across which the
cross-correlation is to be applied. Second, the probability of false
matching needs to be minimized for features with sizes smaller than the
typical displacement between the two epochs.

These two requirements can be met by multiscale cross-correlation
(MCC), which combines the structural and positional information
contained in $S_j$ at all scales of the wavelet decomposition:
\begin{enumerate}
\itemsep0em 
\item The largest scale $J$ of a wavelet decomposition is chosen such
  that the largest expected displacement is smaller than
  $2^J$.

\item Displacements of SSP features are determined at the largest
  scale $J$. For this calculation, all $\Delta^{J + 1}_{J, a}$ are
  set to zero, and $\Delta_{J, a} = \delta_{J, a}$ is calculated for
  each SSP.

\item At each subsequent scale $j$ ($j<J$), $\Delta^{j + 1}_{j, a}$
  is determined first by adopting the displacement $\Delta_{J, a}$
  measured at the $j + 1$ scale for the SSP in which the given
  $j$-scale region $s_{j, a}$ falls. Then the total displacement
  for this SSP is given by $\Delta_{j, a} = \Delta^{j + 1}_{j, a}
  + \delta_{j, a}$.
\end{enumerate}
In this algorithm, the only quantity that needs to be calculated at
each scale is the relative displacement $\delta_{j, a}$. This quantity
can be determined reliably from the cross-correlation.

\section{Applications to astronomical images}
\label{sc:applications}

\begin{figure*}[ht!]
  \centering
  \includegraphics{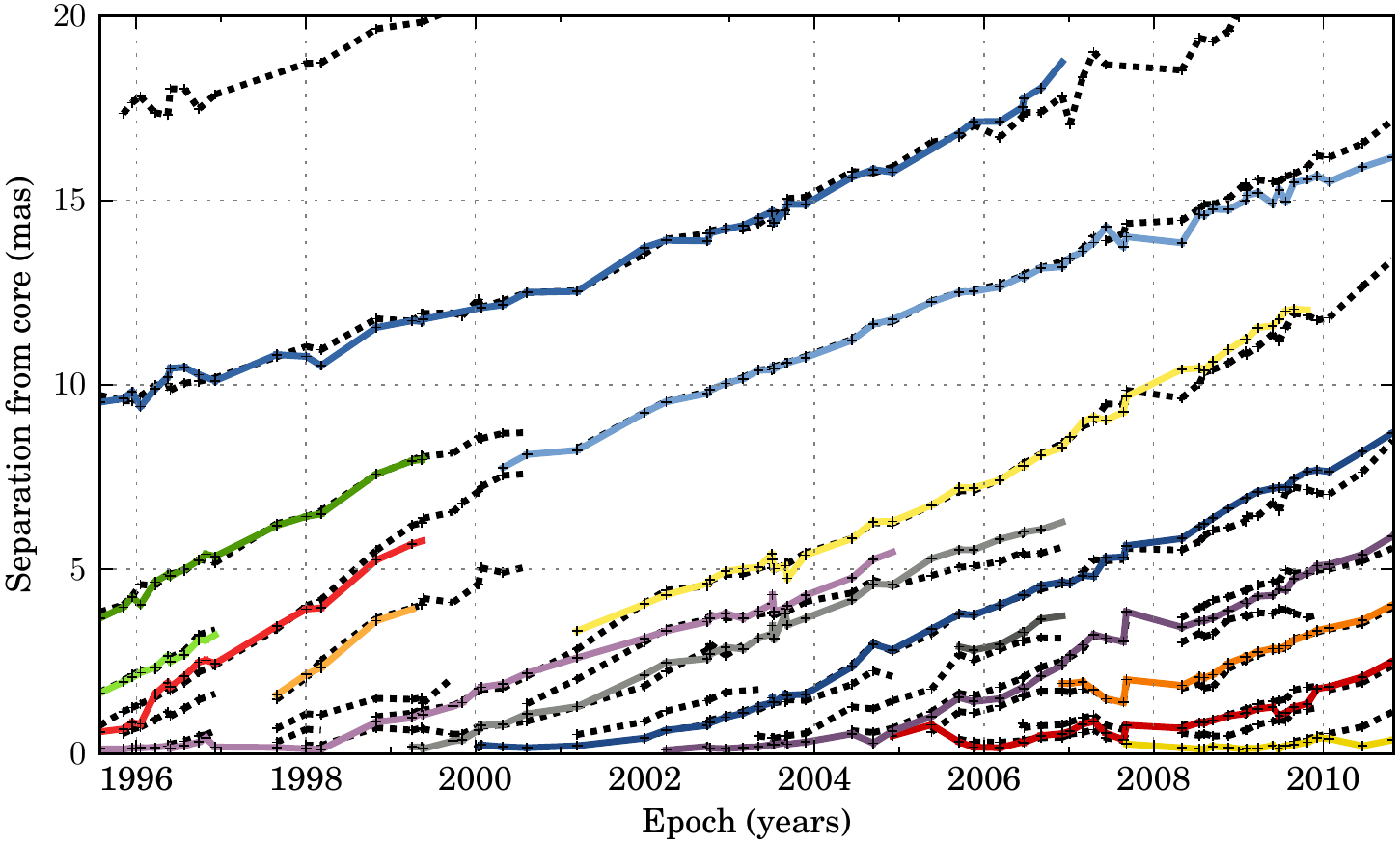}
  \caption{\label{fig:3c273_mojave_dfc_final} Core separation plot of
    the most prominent features in the jet of 3C\,273
    (taken from \cite{mertens_waveletbased_2014}). The model-fit
    based MOJAVE results (dashed lines) are compared with the WISE
    results (solid lines).}
\end{figure*}

We have tested the performance of WISE on astronomical images by
applying it to several image sequences obtained as part of the MOJAVE
long-term monitoring program of extragalactic jets with very long
baseline interferometry (VLBI) observations \citep[][and references
therein]{lister_mojave_2013} and in the M\,87 ``movie'' project
\citep{walker_vlba_2008}. 

 


The analysis of the jet in 3C\,273 was performed on 69 images, with
the observations covering the time range from 1996 to 2010 and providing, on
average, one observation every three months. The core
separations of individual SSP obtained from WISE
decomposition are compared in Fig.~\ref{fig:3c273_mojave_dfc_final})
with the results from the MOJAVE kinematic analysis based on the
Gaussian model fitting of the jet structure.
Comparison indicates that WISE detects
consistently nearly all the components identified by the MOJAVE model
fitting analysis, with a very good agreement on their positional
locations and separation speeds. 

\begin{figure*}
  \centering
  \includegraphics{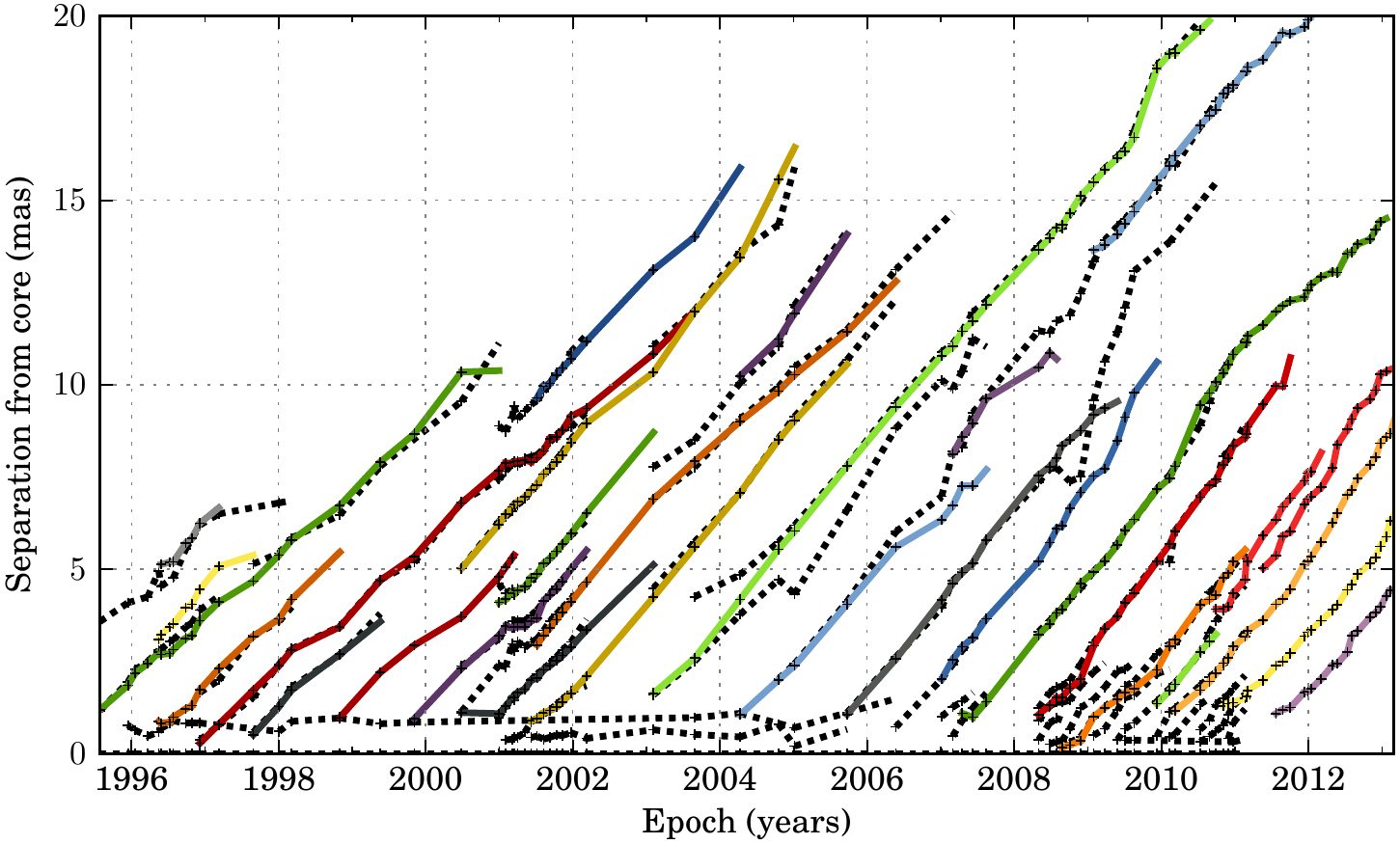}
  \caption{\label{fig:3c120_mojave_dfc_final}
    Core-separation plot of the features identified in the jet of 3C\,120
    (taken from \cite{mertens_waveletbased_2014}).
    The model-fit based MOJAVE results (dashed lines) are compared with the WISE
    results (solid lines).}
\end{figure*}

Similar analysis was performed on 87 observations of the jet in 3C\,120 between
1996 and 2010. A total of 30 moving SSP was detected. The resulting core separations of the
SSP plotted in Fig.~\ref{fig:3c120_mojave_dfc_final} generally agree very well
with the separations of the jet components identified in the MOJAVE Gaussian
model fit analysis.

\begin{figure}
\centering
\includegraphics{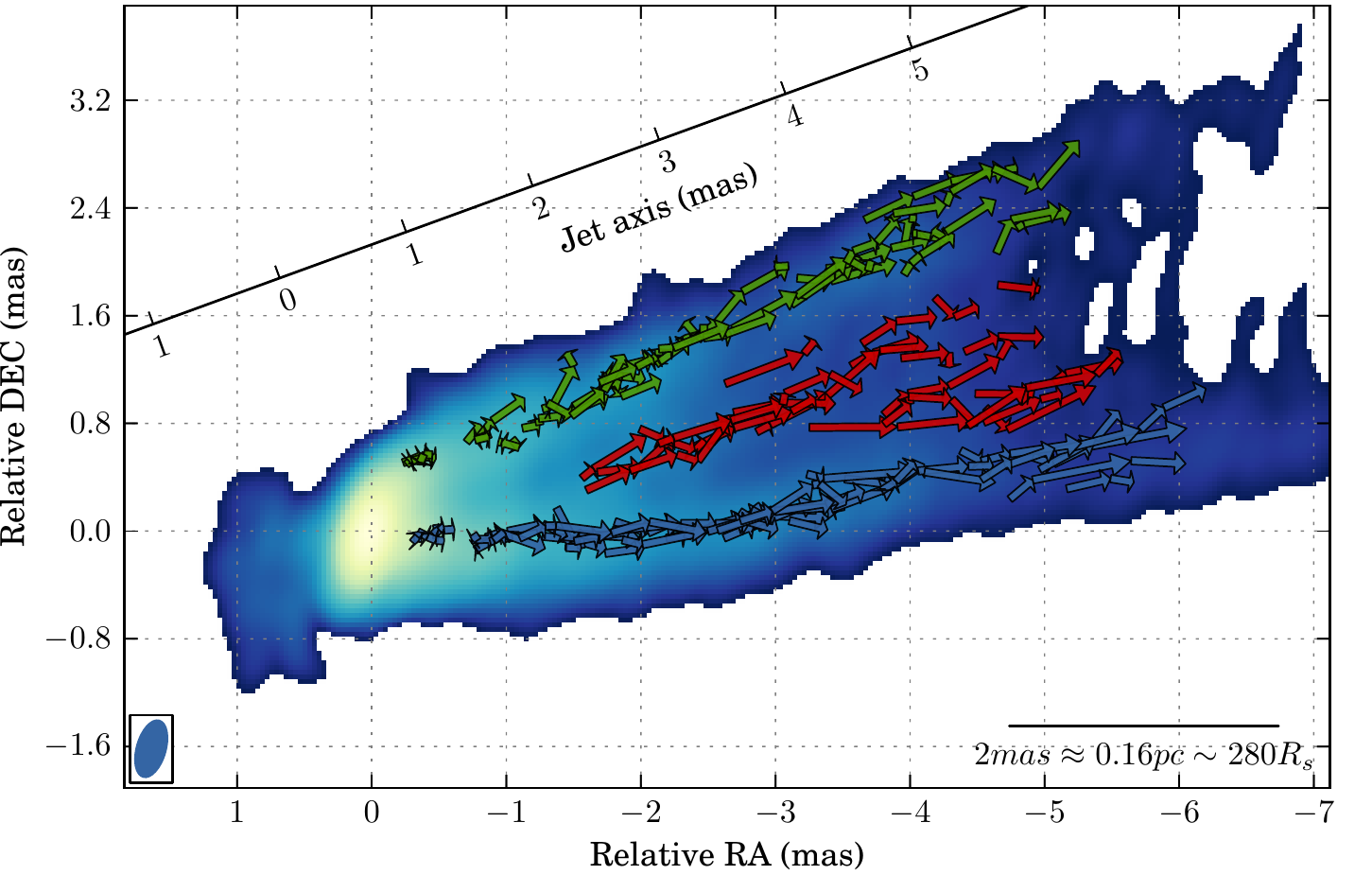}
    \caption{Velocity field in the jet of M\,87. Three main regions are detected:
    a southern (blue) and northern (green) sheath and a central spine (red).
    Arrows represents displacement of individal SSPs between two subsequent
    epochs.}
    \label{fig:m87_velocity_field}
\end{figure}

The jet in M\,87 is probably one of the most investigated. Its proximity
($D=16.7$ Mpc; \cite{jordan_acs_2005}) combined with a large mass of the central
black hole ($M_{BH} \simeq 6.1 \times 10^9 M_{\odot}$;
\cite{gebhardt_blackhole_2011}) make it one of the primary source to probe the
jet formation and acceleration at its base. We analyzed eleven epochs of the
M\,87 jet observed with the Very Large Baseline
Array (VLBA) between 27 January 2007 and 26 August 2007 with an average of 21
days interval \citep{walker_vlba_2008}.  The velocity field obtained from WISE analysis is plotted in
Fig.\ref{fig:m87_velocity_field}. It reveals complex kinematic with
subluminal and superluminal motion measured in three regions of the jet: a
southern and northern sheath and a central spine.

A Stacked Normalized Cross Correlation analysis combining all SSPs obtained at
all eleven epochs was performed. Result indicates the presence of 2 main speed
component in all three regions of the jet. A slow mildly relativistic speed ($
\sim 0.5$ c), and a faster relativistic speed ($\sim 2.25$ c). This could be
explained by the presence of a disk wind around the jet from which only a small
fraction of it at the interface
could be observable through synchrotron radiation \citep{gracia_synthetic_2009}.

Additionally we found that the average velocity measured in the southern sheath
was consistently faster than the average velocity measured in the northern
sheath, suggesting jet or pattern rotation.


\section{Conclusions}
\label{sc:discussion}

The WISE method we presented here offers an effective and
objective way to classify structural patterns in images of
astronomical objects and track their evolution traced by multiple
observations of the same object. The method combines automatic
segmented wavelet decomposition with a multiscale cross-correlation
algorithm, which enables reliable identification and tracking of
statistically significant structural patterns.

We performed the analysis of several transversely resolved AGN jets observed as
part of the MOJAVE project using this method. For all this sources, global
kinematic changes obtained using our SWD technique were in excellent agreement
with the one obtained by the MOJAVE team using model-fitting technique,
demonstrating the reliability of our method for the reconstruction of the
velocity field in transversely resolved flows.

Preliminary results of the application of WISE on the jets of M\,87 suggests a
layered jet with a fast spine surrounded by a subrelativistic disk wind.
Ongoing analysis will concentrate on attempting to model velocity in
combination with the collimation profile, provide an hint on the potential jet
rotation, and investigate the velocities obtained in the counter jet side.


\providecommand{\href}[2]{#2}\begingroup\raggedright\endgroup

\end{document}